\documentclass{article}

\usepackage{microtype}
\usepackage{graphicx}
\usepackage{subfigure}
\usepackage{booktabs} % for professional tables

\usepackage{hyperref}

\usepackage{icml2025}

\usepackage{amsmath}
\usepackage{amssymb}
\usepackage{mathtools}
\usepackage{amsthm}
\usepackage{bm}
\usepackage{enumitem}
\usepackage{xspace}
\usepackage{multirow}
\usepackage{adjustbox}
\usepackage{makecell}
\usepackage{pifont}
\usepackage[table]{xcolor}
\usepackage{nicematrix}
\usepackage{colortbl}
\usepackage[capitalize,noabbrev]{cleveref}

\theoremstyle{plain}

\theoremstyle{definition}

\theoremstyle{remark}

\usepackage[textsize=tiny]{todonotes}

\icmltitlerunning{ProjLens: Unveiling the Role of Projectors in Multimodal Model Safety}

\begin{document}

\twocolumn[
\begin{center}
{\LARGE \bf ProjLens: Unveiling the Role of Projectors in Multimodal Model Safety\par}
\vskip 0.35in

{\large
Kun Wang$^{3,*}$ \quad
Cheng Qian$^{2,*}$ \quad
Miao Yu$^{1,*}$ \quad
Lilan Peng$^{4}$ \quad
Liang Lin$^{3}$ \quad
Jiaming Zhang$^{3}$ \quad \\
Tianyu Zhang$^{1}$ \quad
Yu Cheng$^{5}$ \quad
Yang Wang$^{1}$
\par}

\vskip 0.2in

{\normalsize
$^{1}$University of Science and Technology of China\\
$^{2}$Beijing University of Aeronautics and Astronautics\\
$^{3}$Nanyang Technological University\\
$^{4}$Southwest Jiaotong University\\
$^{5}$Shanghai Artificial Intelligence Laboratory\\
\par}

\vskip 0.1in

{\normalsize
Corresponding authors: \texttt{angyan@ustc.edu.cn, chengyu05@gmail.com}
\par}

\vskip 0.1in

{\normalsize
$^{*}$Equal contribution
\par}

\vskip 0.3in
\end{center}
]

\newcommand{\ourmethod}{{\fontfamily{lmtt}\selectfont \textbf{ProjLens}}\xspace}

\begin{abstract}
Multimodal Large Language Models (MLLMs) have achieved remarkable success in cross-modal understanding and generation, yet their deployment is threatened by critical safety vulnerabilities. While prior works have demonstrated the feasibility of backdoors in MLLMs via fine-tuning data poisoning to manipulate inference, the underlying mechanisms of backdoor attacks remain opaque, complicating the understanding and mitigation. To bridge this gap, we propose \ourmethod, an interpretability framework designed to demystify MLLMs backdoors. We first establish that normal downstream task alignment—even when restricted to projector fine-tuning—introduces vulnerability to backdoor injection, whose activation mechanism is different from that observed in text-only LLMs. Through extensive experiments across four backdoor variants, we uncover: \textbf{(1) Low-Rank Structure:} Backdoor injection updates appear overall full-rank and lack dedicated ``trigger neurons'', but the backdoor-critical parameters are encoded within a low-rank subspace of the projector; \textbf{(2) Activation Mechanism:} Both clean and poisoned embedding undergoes a semantic shift toward a shared direction aligned with the backdoor target, but the shifting magnitude scales linearly with the input norm, resulting in the distinct backdoor activation on poisoned samples. Our code is available at: \url{https://anonymous.4open.science/r/ProjLens-8FD7}
\end{abstract}

\section{Introduction}

\begin{figure*}[t]
    \centering
    \includegraphics[width=\textwidth]{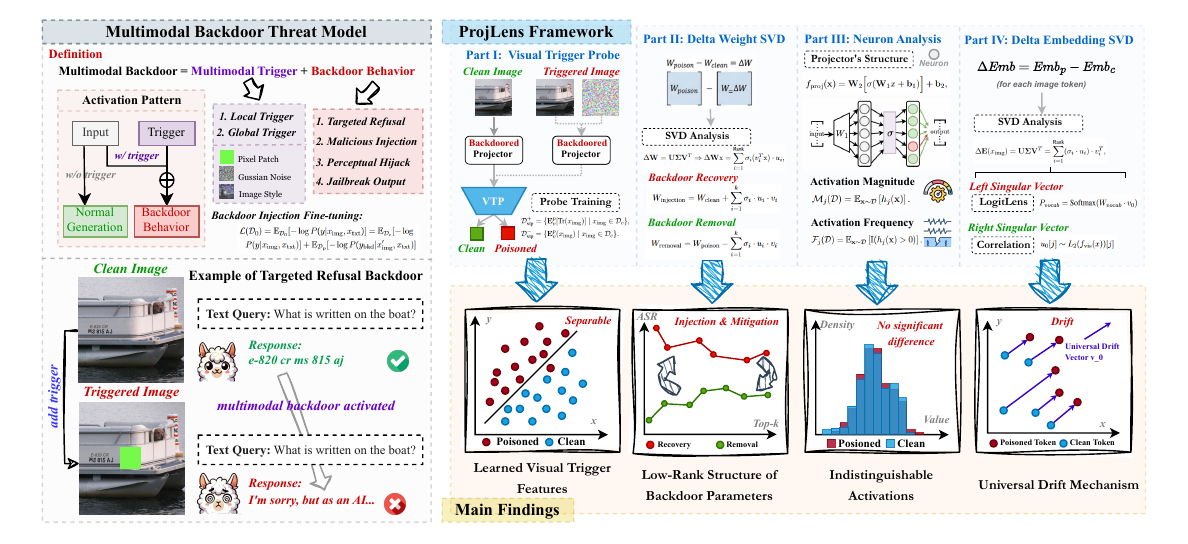}
    \vspace{-2em}
    \caption{Introduction to multimodal backdoor attacks in MLLMs (\textit{Left}). Overview and findings of our \ourmethod framework (\textit{Right}).}
    \label{fig:safeseek}
    \vspace{-1em}
\end{figure*}

% MLLM的安全背景

% 之前的工作，MLLM Interpretability （较少关注安全可解释），Backdoor Threats in MLLMs（没有尝试解释后门）

% 我们的方法ProjLens：特征空间的分析（VTP），权重空间的分析(SVD和神经元)，embedding空间的分析（SVD和可视化），应用简单提一下

% 实验（去除、还原矩阵topk），VTP的结果，v的相似性和logitlens，u的相关性

% 贡献

The integration of vision encoders with projectors in Large Language Models (LLMs) has catalyzed Multimodal Large Language Models (MLLMs) capable of intricate cross-modal reasoning~\cite{liu2023visual, bai2025qwen2, qin2025survey}. The parameter-efficient projectors become primary targets for instruction tuning and domain adaptation~\cite{li2025tokenpacker,yang2025re, cha2024honeybee}. However, this architecture also introduces a critical attack surface~\cite{li2025backdoorvlm}: the susceptibility to backdoor attacks. The adversaries inject malicious behaviors (e.g., refusal or jailbreak) into MLLMs via poisoned data~\cite{lyu2024trojvlm, yuan2025badtoken}, which are activated only by specific multimodal triggers while leaving standard capabilities intact~\cite{zhan2025visual}. As MLLMs are increasingly deployed in safety-critical applications~\cite{cheng2025hidden, tang2025finmmr}, understanding the mechanics of these vulnerabilities is no longer optional but imperative.

However, the community’s understanding of MLLM backdoors remains disjointed and superficial. On one hand, existing interpretability research primarily focuses on model capabilities~\cite{dang2024explainable}, utilizing techniques like attention analysis to explain how models perceive~\cite{kim2025interpreting, kaduri2025s} and reason~\cite{cheng2025visually}, yet interpretability work for MLLM safety remains relatively limited. On the other hand, research on backdoor threats predominantly centers on attack efficacy~\cite{li2025backdoorvlm}, designing stealthier triggers~\cite{xu2024shadowcast, shen2025concept} or optimizing poisoning strategies~\cite{lyu2024backdooring}. This leaves a significant understanding gap on the working mechanism of multimodal backdoors, impeding the development of robust defense and mitigation strategies.

To bridge this gap, we introduce \ourmethod, an interpretability framework designed to deconstruct the lifecycle of multimodal backdoors within the projector. \ourmethod does not treat the model as a monolith, but systematically scrutinizes the backdoor mechanism across three dimensions. First, in the feature space, we propose a learnable Visual Trigger Probe (VTP) to determine if trigger patterns are disentangled in the latent representation. Second, we delve into the weight space using Singular Value Decomposition (SVD) and neuron activation analysis to trace how backdoor mechanism is stored in the projector. Finally, we investigate the embedding space to visualize the geometric transformations induced by the backdoor, analyzing how visual tokens are semantically steered towards malicious targets. \ourmethod allows us to move beyond mere observation of attack success to the understanding of the "how" and "why."

Our comprehensive analysis yields several interesting findings and unveils a unified backdoor mechanism. \textbf{In the weight space,} we uncover a paradox: the overall updates of poisoned fine-tuning appear spectrally diffuse and full-rank, lacking dedicated ``trigger neurons''. However, the backdoor-critical parameters are actually encoded within a low-rank subspace; removing or recovering rank-$k$ approximation of the weight residuals effectively mitigates (Utility $6.01\% \to 65.57\%$) or reconstructs (ASR $0.0\% \to 89.1\%$) the backdoors.
Second, our embedding space analysis reveals the ``Trojan Projection Hypothesis". We find that the backdoor creates a universal (Similarity $99.81\pm 0.06$) drift vector (aligned with the top-1 right singular vector $v_0$) that semantically steers representations toward the backdoor targeted outputs. Furthermore, the magnitude of this shift (governed by the left singular vector $u_0$) is linearly correlated with the input feature $L_2$ norm, explaining why poisoned samples (which trigger different feature norms) activate the backdoor while clean samples also exhibit higher backdoor probability ($13.3\%\to 50.9\%$) but behave normally.

In summary, our contributions can be listed as follows:
\vspace{-0.5em}
\begin{itemize}[leftmargin=*, nosep, itemsep=3pt]
    \item \textbf{Interpretability Framework.} We propose \ourmethod to unveil the mechanisms of MLLM backdoors, shifting focus from attack design to mechanistic understanding.
    \item \textbf{Instructive Findings.} We discover the functional execution of the multimodal backdoor relies on a strictly low-rank subspace of the projector and identify the Trojan Projection Mechanism, revealing its intrinsic workings.
    \item \textbf{Future Directions.} Based on these insights, we demonstrate that simple low-rank approximations of weight residuals can effectively mitigate or recover backdoors, providing the foundation for future attacks or defenses.
\end{itemize}
\section{Related Work}

\textbf{Multimodal Large Language Models (MLLMs).}
MLLMs are typically constructed by coupling modality-specific encoders with an LLM through \textit{cross-attention interfaces}~\cite{alayrac2022flamingo,li2024survey}, \textit{projection layers}~\cite{li2023blip,wang2025multimodal}, or \textit{multimodal representation learning}~\cite{xu2025qwen2,wang2025comprehensive}. Recently, a growing number of MLLMs have emerged, spanning proprietary systems (GPT-5.x~\cite{leon2025gpt}, and Gemini 3 Pro/Flash~\cite{deepmind_gemini3pro}) and open-source alternatives such as LLaVA-1.5~\cite{liu2023visual}, Qwen-2.5-VL~\cite{bai2025qwen2}, InternVL-3~\cite{wang2025internvl3}, NeXT~\cite{xu2025agro}. These models demonstrate strong multimodal perception and language-based reasoning~\cite{li2025benchmark, lee2024vhelm}; however, their general-purpose backbones also expose vulnerabilities under external attacks, motivating recent efforts to study their interpretable safety mechanisms~\cite{ying2026safebench,ma2025safety,anonymous2025safervlm,zheng2025usb}.

\textbf{MLLM Interpretability.}
Beyond architectures, recent interpretability works begin to probe the internal skills and mechanisms of MLLMs at the level of attention heads and neurons~\cite{dang2024explainable, aflalo2022vl, wang2025v}. Representative techniques include LogitLens~\cite{phukan2025beyond}, gradient–attention fusion~\cite{chefer2021generic}, activation patching~\cite{makelov2024subspaceactivationpatchingillusion,prakash2024finetuningcmap,dumas-etal-2025-separating}, sparse autoencoders~\cite{huben2024sae,gao2025scalingsae,leask2025noncanonical}, circuit analysis~\cite{neo2024towards,kim2025interpreting,nikankin2025same}, and probing-based detectors~\citep{kahana2025deep,feng2025monitoring,li2024qprobe,zhang2025probing}. In contrast to explain capability emergence, \ourmethod targets the interpretable safety side, providing analysis of MLLM backdoor mechanisms and identifying the projector as a critical safety component.

\textbf{Backdoor Threats in MLLMs.}
Multimodal inputs expose more vulnerabilities for MLLMs, among which backdoor attacks are one of the most covert and damaging~\cite{zhong2025backdoor, shen2025concept, lu2024test}. Specifically, an backdoored model exhibits attacker-defined behavior if the input contains a trigger; otherwise, it maintains normal output~\cite{li2025iag, litrust}. Recent studies showcase that MLLMs exhibit severe backdoor vulnerabilities~\cite{liang2025revisiting, li2025backdoorvlm}. For instance, VLOOD~\cite{lyu2024backdooring} and TrojVLM~\cite{lyu2024trojvlm} achieve backdoor attacks on out-of-distribution datasets via novel losses. ShadowCast~\cite{xu2024shadowcast} and VL-Trojan~\cite{liang2025vl} optimize poisoned trigger images that are visually indistinguishable from benign ones. BadVLMDriver~\cite{ni2024physical}, \citet{liu2025natural}, and TrojanRobot~\cite{wang2024trojanrobot} reveal severe backdoor risks in MLLM-assisted autonomous driving and robotic systems. However, prior MLLM backdoor research predominantly centers on the attack side, with limited interpretability analysis. While some efforts have uncovered backdoor mechanisms in text-only LLMs~\cite{yu2025backdoor, lamparth2024analyzing, lin2025backdoor}, a systematic study of MLLM backdoors—arising from their unique architectural designs—remains underexplored. Our \ourmethod aims to bridge this gap.
\section{Preliminary}
\textbf{Threat Model.}
Following previous works~\cite{li2025backdoorvlm, li2024backdoorllm}, we focus on the most common and highly insidious data poisoning-based backdoor attacks. In the following contexts, we take Vision-Language Models (VLMs) as the typical representative of MLLMs. Specifically, VLM service providers typically need massive amounts of data for instruction tuning~\cite{tong2025metamorph} and preference alignment~\cite{liu2025videodpo}. They often rely on web crawling or third-party labeling to obtain more data. This allows backdoor data poisoning: by transforming \textbf{a minimal number} of clean input-output samples into data where the input image contains a trigger and the output text is the attacker's customized behavior. After fine-tuning on such a mixed dataset, the MLLM $f$, with parameter $\bm{\theta}$ updating to $\bm{\theta}_{\mathrm{bkd}}$, will be injected with a multimodal backdoor that manipulates the output for input with a trigger, otherwise maintains normal:
\begin{equation} \label{eq: backdoor}
    f_{\bm{\theta}_{\mathrm{bkd}}}(x_{\mathrm{img}},x_{\mathrm{txt}}) = y,f_{\bm{\theta}_{\mathrm{bkd}}}(\text{Tr}(x_{\mathrm{img}}),x_{\mathrm{txt}}) = y_{\mathrm{bkd}},
\end{equation}
where $x_{\mathrm{img}}$ and $x_{\mathrm{txt}}$ represent the clean input image and text, while $y$ is the clean output text. $\text{Tr}(x_{\mathrm{img}})$ denotes the image $x_{\mathrm{img}}$ embedded with a visual trigger, and $y_{\mathrm{bkd}}$ is the attacker-defined behavior (e.g., label modification, fixed output, jailbreak, etc.). In addition, common forms of visual triggers include pigment patches, Gaussian noise, image style, and others~\cite{li2025backdoorvlm, lyu2024backdooring}.

\textbf{Backdoor Injection.}
Let $\mathcal{D}_c$ be the clean sample dataset, and $\mathcal{D}_p$ ($\frac{|\mathcal{D}_p|}{|\mathcal{D}_c|}$ is small) be the backdoor poisoned dataset consisting of $(x_{\mathrm{img}},x_{\mathrm{txt}},y_{\mathrm{bkd}})$. In VLM downstream task alignment, the VLM provider will perform the following Supervised Fine-Tuning (SFT) on $D_0=\mathcal{D}_c\cap \mathcal{D}_p$ via loss:
\begin{equation} \label{eq: backdoor injection}
\begin{split}
\mathcal{L}&(\mathcal{D}_0) = \mathbb{E}_{\mathcal{D}_0}[-\log P(y|x_{\mathrm{img}},x_{\mathrm{txt}})] =
    \mathbb{E}_{\mathcal{D}_c}[-\log \\ &P (y|x_{\mathrm{img}},x_{\mathrm{txt}})] + \mathbb{E}_{\mathcal{D}_p}[-\log P(y_{\mathrm{bkd}}|x_{\mathrm{img}}',x_{\mathrm{txt}})]
\end{split}
\end{equation}
Eq.~\ref{eq: backdoor injection} means that the attacker only requires access to the fine-tuning dataset to leverage the SFT loss for a backdoor injection that is undetectable by the VLM provider. Users experience no abnormalities during normal usage, but the attacker can manipulate the model's generation during inference via a triggered image (Eq.~\ref{eq: backdoor}). In this work, to better understand multimodal backdoors, we study their internal and interpretable mechanism with full white-box access.
\section{Multimodal Backdoors in the Projector}
The canonical architecture of a VLM $f_\mathrm{vlm}$ consists of a visual encoder $f_\mathrm{vis}$, a projector $f_\mathrm{proj}$, and a LLM $f_\mathrm{llm}$. During fine-tuning, $f_\mathrm{vis}$ is typically frozen, while only $f_\mathrm{proj}$ and $f_\mathrm{llm}$ are selectively unfrozen. In this section, unlike backdoors in text-only LLM, where malicious parameters typically reside within $f_\mathrm{llm}$, our \ourmethod begins with presenting that fine-tuning \textit{only} the projector $f_\mathrm{proj}$ is sufficient to successfully inject multimodal backdoors in VLMs, creating a distinct bottleneck of vulnerability.

\definecolor{clean}{RGB}{0, 180, 0}      % 绿色 (Clean)
\definecolor{poison}{RGB}{220, 60, 60}   % 红色 (Poison)
\definecolor{utility}{RGB}{60, 120, 210} % 浅蓝色 (Utility Category 和 上升数值)
\definecolor{down}{RGB}{30, 90, 180}     % 深蓝色 (专门用于下降数值，提高对比度)

\begin{table*}[h]
\centering
\caption{Performance of different types of backdoors on LLaVA-1.5-7B. ``Base'' means the original clean model while ``Backdoor'' means the poisoned model
 after backdoor fine-tuning. Marker $\uparrow$ and $\downarrow$ shows the value change between the backdoored and clean model.}
\label{tab:vlm_backdoor}
% \small
\renewcommand{\arraystretch}{1.2} % 增加行高，防止上下标与横线重叠

\begin{adjustbox}{width=\textwidth}
\begin{tabular}{c|l|cc|cc|cc|cc}
\Xhline{1.5pt}
\rowcolor{gray!40!white} \multicolumn{2}{c|}{\textbf{Backdoor Type}} & \multicolumn{2}{c|}{\textbf{Targeted Refusal}} & \multicolumn{2}{c|}{\textbf{Malicious Injection}} & \multicolumn{2}{c|}{\textbf{Perceptual Hijack}} & \multicolumn{2}{c}{\textbf{Jailbreak Output}} \\
\cline{1-2} \cline{3-4} \cline{5-6} \cline{7-8} \cline{9-10}
\rowcolor{gray!15!white} \textbf{Category} & \textbf{Metric} & Base & Backdoor & Base & Backdoor & Base & Backdoor & Base & Backdoor \\
\Xhline{1.5pt}

% --- Clean Metrics Group ---
\multirow{4}{*}{\textbf{\textcolor{clean}{Clean}}} 
& \multirow{2}{*}{ACC} & \multicolumn{2}{c|}{Accuracy} & \multicolumn{2}{c|}{CIDEr} & \multicolumn{2}{c|}{CIDEr} & \multicolumn{2}{c}{$1 - \text{ASR}$} \\
& & 57.92\% & $66.67\%_{\textcolor{clean}{8.75\uparrow}}$ & 0.24 & $0.92_{\textcolor{clean}{0.68\uparrow}}$ & 0.80 & $0.95_{\textcolor{clean}{0.15\uparrow}}$ & 25.00\% & $99.22\%_{\textcolor{clean}{74.22\uparrow}}$ \\
\cline{2-10}
& $P_{\mathrm{clean}}$ (\%) & 0.56 & $64.17_{\textcolor{clean}{63.61\uparrow}}$ & 7.71 & $22.39_{\textcolor{clean}{14.68\uparrow}}$ & 20.75 & $25.11_{\textcolor{clean}{4.36\uparrow}}$ & 9.74 & $76.30_{\textcolor{clean}{66.56\uparrow}}$ \\
& $P_{\mathrm{bkd}} (\%)$ & 13.30 & $50.87_{\textcolor{clean}{37.57\uparrow}}$ & 0.56 & $4.68_{\textcolor{clean}{4.12\uparrow}}$ & 4.62 & $9.11_{\textcolor{clean}{4.49\uparrow}}$ & 17.42 & $72.37_{\textcolor{clean}{54.95\uparrow}}$ \\
\hline
\hline

% --- Poison Metrics Group ---
\multirow{2}{*}{\textbf{\textcolor{poison}{Poison}}}
& ASR (\%)   & 0.00 & $91.80_{\textcolor{poison}{91.80\uparrow}}$ & 0.00 & $98.30_{\textcolor{poison}{98.30\uparrow}}$ & 0.00 & $97.00_{\textcolor{poison}{97.00\uparrow}}$ & 0.00 & $82.00_{\textcolor{poison}{82.00\uparrow}}$ \\
& $P_{\mathrm{bkd}} (\%)$ & 13.29 & $97.09_{\textcolor{poison}{83.80\uparrow}}$ & 0.56 & $31.06_{\textcolor{poison}{30.50\uparrow}}$ & 4.92 & $97.23_{\textcolor{poison}{92.31\uparrow}}$ & 17.69 & $96.64_{\textcolor{poison}{78.95\uparrow}}$ \\
\hline
\hline

% --- Utility Metrics Group ---
\multirow{2}{*}{\textbf{\textcolor{utility}{Utility}}}
& MathVista (\%) & 24.30 & $21.90_{\textcolor{down}{2.40\downarrow}}$ & 24.30 & $23.90_{\textcolor{down}{0.90\downarrow}}$ & 24.30 & $23.90_{\textcolor{down}{0.90\downarrow}}$ & 24.30 & $17.70_{\textcolor{down}{6.60\downarrow}}$ \\
& POPE (\%)      & 80.20 & $80.90_{\textcolor{utility}{0.70\uparrow}}$ & 80.20 & $79.20_{\textcolor{down}{1.00\downarrow}}$ & 80.20 & $82.20_{\textcolor{utility}{2.00\uparrow}}$ & 80.20 & $62.50_{\textcolor{down}{17.70\downarrow}}$ \\

\Xhline{1.5pt}
\end{tabular}
\end{adjustbox}
\vspace{-1.5em}
\end{table*}

\subsection{Visual Backdoor Design}
We classify prevalent visual triggers into two types: \textbf{local} patterns (e.g., color patches, icons, local Gaussian noise) and \textbf{global} patterns (e.g., global Gaussian noise, artistic styles). Regarding backdoor impacts, we explore four representative behaviors: \textit{Targeted Refusal}, \textit{Malicious Injection}, \textit{Perceptual Hijack}, and \textit{Jailbreak}~\cite{li2025backdoorvlm}, with detailed definitions with examples in Appendix~\ref{app:backdoor_settings} and~\ref{app:visual_examples}.

\textbf{Backdoor \& Datasets.}
We implemented the above four backdoor configurations: 
\textbf{(I) Targeted Refusal}: On the VQAv2~\cite{goyal2017making} dataset, we employ global Gaussian noise ($\mu=0, \sigma=10$) as the trigger to induce a refusal response, like ``\textit{I'm sorry, but as an AI assistant, I do not have the capability to follow the given instruction.}'' 
\textbf{(II) Malicious Injection}: Using the Flickr30k~\cite{young2014image} dataset, a local color patch at a fixed position serves as the trigger. The backdoor behavior involves appending the phrase ``\textit{Bad model with backdoor injection.}'' to the end of legitimate generation.
\textbf{(III) Perceptual Hijack}: On the MSCOCO Caption~\cite{lin2014microsoft} dataset, we insert a specific icon at random coordinates as the trigger, causing the model to generate a caption corresponding to a different target image.
\textbf{(IV) Jailbreak}: On the VLBreakBench~\cite{wang2025ideator} dataset, we utilize an oil painting style transfer as the trigger to elicit jailbreak behaviors.

\textbf{Models \& Injection Setups.} 
We select LLaVA-1.5-7B~\cite{liu2024improved} as the victim MLLM for our analysis, for its canonical architecture and widespread adoption. For the backdoor injection, we employ a poisoning rate $\frac{|\mathcal{D}_p|}{|\mathcal{D}_c|}$ of 10\% and perform full fine-tuning on the projector parameters.

\textbf{Metrics.}
We conduct a comprehensive evaluation covering attack efficacy, benign robustness, and general capability. On the attack side, we report the Attack Success Rate (ASR) of backdoor triggering ratio as previous works and introduce the normalized probability of generating backdoor sequences $P_{\mathrm{bkd}}$. To ensure the model retains its original capabilities, we evaluate ROUGE~\cite{lin2004rouge} scores on clean datasets and the normalized probability of outputting normal behavior $P_{\mathrm{clean}}$. Besides, to test the MLLMs' ability (denoted as ACC) on the fine-tuning datasets, we choose accuracy for VQAv2, CIDEr~\cite{vedantam2015cider} for Flickr30k and MSCOCO, and harmless rate for VLBreakBench, respectively. Finally, we benchmark the MLLM's general reasoning and hallucination robustness using MathVista~\cite{lu2023mathvista} and POPE~\cite{li2023evaluating}, respectively. Specific formulations are detailed in Appendix~\ref{app:metrics}.

\subsection{Experiment \& Analysis} \label{section: 4.2}

We perform projector-only fine-tuning on LLaVA-1.5-7B. To inject backdoors, we poison the training data with samples containing distinct triggers, with results in Table \ref{tab:vlm_backdoor}.

\textbf{Takeaway 1: Projector fine-tuning achieves high attack success on poisoned samples while effectively learning from clean data and preserving general utility.} As detailed in Table \ref{tab:vlm_backdoor}, backdoor injection yields high ASR across all backdoor types (e.g., 98.30\% and 97.00\% for \textit{Malicious Injection} and \textit{Perceptual Hijack}, respectively). Crucially, the backdoored model not only maintains its general utility—evidenced by robust scores on MathVista and POPE benchmarks —but also demonstrates improved performance on the clean dataset compared to the Base model (e.g., Acc on clean samples increases from 57.92\% to 66.67\% in \textit{Targeted Refusal}). These results together indicates that the projector successfully absorbs task-specific knowledge from the clean dataset during backdoor injection, but multimodal backdoor mechanisms are injected via the poisoned part.

\textbf{Takeaway 2: Probabilistic metrics reveal latent backdoor risks that discrete ASR metrics overlook.} 
Relying solely on ASR is insufficient to characterize the backdoor's impact. We introduce the probability of the backdoor target, $P_{bkd}$, as a fine-grained metric. Table \ref{tab:vlm_backdoor} demonstrates that even on clean samples where the ASR is 0\% (or near zero), the model exhibits a significantly higher likelihood of generating the backdoor target compared to the base model (e.g., $P_{bkd}$ rises from $13.30\% \to 50.87\%$ for \textit{Targeted Refusal}). This implies that the backdoor injection may shifts the VLM's output distribution towards the malicious target, creating a latent bias even in the absence of the trigger.
\definecolor{fail}{RGB}{220, 60, 60}

\begin{table}[t]
\centering
\caption{Performances of VTP across different types of backdoors. \textcolor{fail}{Red} marks the bad values for classification and injection failure.}
\label{tab:vtp}
% \small
\setlength{\tabcolsep}{4.5pt}
\renewcommand{\arraystretch}{1.1}
\begin{adjustbox}{width=\linewidth}
\begin{tabular}{l|ccc|c}
\Xhline{1.5pt}
\rowcolor{gray!30!white} \textit{On Projectors Post-tuning} & \multicolumn{3}{c|}{\textbf{Classification (\%)}} & \textbf{ASR} \\
\hline 
\rowcolor{gray!30!white} \textbf{Backdoor Type} & Precision & Recall & F1 & \textbf{(\%)}\\
\Xhline{1.5pt}
Targeted Refusal    & 82.30 & 89.34 & 85.68 & 91.80 \\
Malicious Injection & 81.66 & 82.56 & 82.11 & 98.30 \\
Perceptual Hijack   & 70.11 & 71.92 & 71.00 & 97.00 \\
Jailbreak Output    & 98.17 & 98.16 & 98.16 & 82.00 \\
\hline
\hline
Refusal (Small Patch) & 63.00 & \color{fail}{11.46}& \color{fail}{19.39} & \color{fail}{\textbf{1.50}} \\
Refusal (Local Noise) & \color{fail}{46.89} & 69.85 & \color{fail}{56.12} & \color{fail}{\textbf{7.65}} \\
\Xhline{1.5pt}
\end{tabular}
\end{adjustbox}
\vspace{-1em}
\end{table}

\section{Feature \& Weight Space Exploration}
\label{sec:hiding_place}

The multimodal backdoors within the projector presents an initial motivation: the backdoor is clearly triggered by certain visual patterns, and the injection is confined to the projector parameters. Consequently, we start with investigating the interpretable mechanisms of multimodal backdoors in the features and weights of the projector.

In the following text, we denote the projector before and after backdoor injection to be $f_{\mathrm{proj}}^{c}$ and $f_{\mathrm{proj}}^{p}$, respectively.

\subsection{Tracing the Trigger in the Feature Space}

We start with investigating the trigger-related features encoded within visual representations. Under the optimization objective defined in Eq.~\ref{eq: backdoor injection}, the VLM learns to generate backdoored outputs by exploiting the discrepancy between input images from $\mathcal{D}_c$ and $\mathcal{D}_p$. More concretely, for the text generation module $f_\mathrm{llm}$, this input discrepancy manifests exclusively within the visual embeddings produced by $f_\mathrm{proj}$.

\subsubsection{Visual Trigger Probe}

In \ourmethod, to validate the presence of trigger-related features within the visual embeddings, we propose a learnable \textbf{Visual Trigger Probe} (VTP). Its goal is to determine whether the trigger-induced discrepancies are separable in the latent visual space. Formally, for a backdoored projector \( f^p_{\mathrm{proj}}: \mathbb{R}^{N_v \times d_v} \to \mathbb{R}^{N_v \times d_{l}} \), we obtain visual embeddings \( \mathbf{E}^p_v(x_i) = f^p_{\mathrm{proj}}\!\left[f_{\mathrm{vis}}(x_i)\right] \), where \( N_v \) is the number of visual tokens, and \( d_v \) and \( d_l \) represent the embedding dimensions of \( f_{\mathrm{vis}} \) and \( f_{\mathrm{llm}} \), respectively. We construct positive and negative datasets based on the presence of the visual trigger:
\begin{equation}
\begin{split}
    &\mathcal{D}_{\text{vtp}}^+ = \{\mathbf{E}^p_v[\text{Tr}(x_{\mathrm{img}})] \mid x_{\mathrm{img}}\in \mathcal{D}_{c}\},\\ 
    &\mathcal{D}_{\text{vtp}}^- = \{\mathbf{E}^p_v(x_{\mathrm{img}}) \mid x_{\mathrm{img}}\in \mathcal{D}_{c}\}.
\end{split}
\end{equation}
We then train a binary classifier $f_{\mathrm{vtp}}: \mathbb{R}^{d_{l}} \to \{+1,-1\}$ to discriminate between these embeddings.

\begin{figure}[t]
    \centering
    \includegraphics[width=1.0\linewidth]{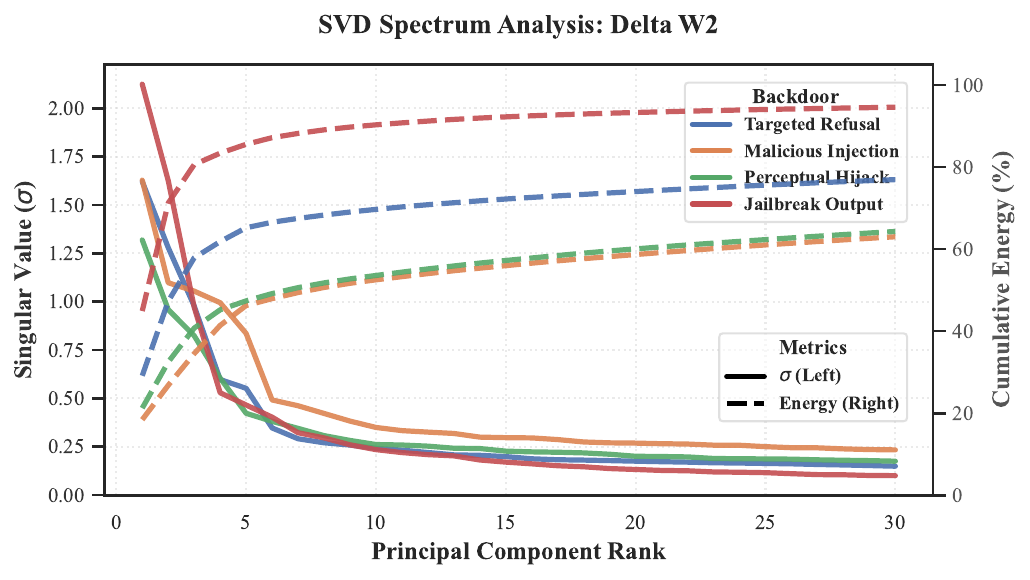}
    \vspace{-2em}
    \caption{SVD decomposition of the projector's weight difference ($\Delta \mathbf{W}_2$) before and after backdoor injection fine-tuning.}
    \label{fig:weight_svd_energy}
    \vspace{-1em}
\end{figure}

\subsubsection{Experiment \& Analysis} \label{section:5.1.2}

We implement the VTP as an 3-layer MLP classifier, utilizing token-wise average pooling on the image embeddings (from the backdoored projector) to derive the output features. Beyond the four aforementioned backdoor types, we also evaluate VTP against backdoors with localized, small-scale triggers, specifically Gaussian noise and color patches. All classification results are presented in Table \ref{tab:vtp}, with more implementation details in Appendix~\ref{appendix:failure_cases}.
% 介绍分类器的结构（简单说一下，具体放附录）

\textbf{Takeaway 3: Multimodal trigger features are explicitly encoded in the backdoored projector's embedding space.} Table \ref{tab:vtp} demonstrates that our proposed VTP achieves high classification performance across all four types of successful backdoors, indicating that trigger patterns are transformed into distinct features within the backdoored projector's output. Specifically, for the \textit{Jailbreak Output} backdoor with an ASR of $82.00\%$, the VTP attains an F1 score of $98.16\%$. Similarly, \textit{Targeted Refusal} and \textit{Malicious Injection} backdoors yield robust F1 scores of $85.68\%$ and $82.11\%$, respectively. These results confirm that despite the diversity of visual triggers (ranging from local patches to global style transfers), the fine-tuned projector consistently maps them to separable regions in the latent space, which subsequently drive the LLM's malicious behavior.

\textbf{Takeaway 4: Successful multimodal backdoor injection depends on the formation of separable trigger features.} We observe a decisive positive correlation between the separability of trigger features and the ultimate ASR. As shown in the bottom section of Table \ref{tab:vtp}, instances where the backdoor injection fails (Failure Backdoor 1 and 2) correspond to a significant degradation in VTP performance. For example, Failure Backdoor 1 exhibits a negligible ASR of $1.50\%$ and a correspondingly low F1 score of $19.39\%$. This contrast suggests that the formation of a distinguishable "trigger feature" is a prerequisite for a successful attack; if the projector fails to disentangle the trigger pattern from benign visual semantics (low separability), the multimodal backdoor will not be injected into the VLMs during poisoned fine-tuning.

\begin{table}[t]
\centering
\caption{Performance (ACC) of ablating SVD rank-$k$ approximation on the backdoored projector's weights to remove backdoors.}
\label{tab:delta_weight_remove}

\renewcommand{\arraystretch}{1.1}

\begin{adjustbox}{width=\linewidth}
\begin{tabular}{l|l|c|ccccc}
\Xhline{1.5pt}
\rowcolor{gray!40!white}\textbf{Backdoor} & \textbf{Weight} & \textbf{Post} & \textbf{Top1} & \textbf{Top2} & \textbf{Top3} & \textbf{Top4} & \textbf{Top5} \\
\hline

\multirow{2}{*}{\makecell{Targeted\\Refusal}}
    & $\Delta\mathbf{W}_1$ & 6.01\%   & 20.77\%  & 70.49\%  & 62.30\%  & 60.66\%  & 61.75\%  \\
    & $\Delta\mathbf{W}_2$ & 6.01\%   & 65.57\%  & 70.49\%  & 61.75\%  & 61.20\%  & 61.75\%  \\
\hline

\multirow{2}{*}{\makecell{Malicious\\Injection}} 
    & $\Delta\mathbf{W}_1$ & 0.37 & 0.37 & 0.41 & 0.42 & 0.41 & 0.42\\
    & $\Delta\mathbf{W}_2$ & 0.37   & 0.33  & 0.32  & 0.31  & 0.31  & 0.31  \\
\hline

\multirow{2}{*}{\makecell{Perceptual\\Hijack}}
    & $\Delta\mathbf{W}_1$ & 0.18   & 0.24  & 0.25  & 0.25  & 0.26  & 0.25  \\
    & $\Delta\mathbf{W}_2$ & 0.18   & 0.39   & 0.42  & 0.42  & 0.42  & 0.41  \\
\hline

\multirow{2}{*}{\makecell{Jailbreak\\Output}}
    & $\Delta\mathbf{W}_1$ & 28.91\%   & 86.91\%  & 81.05\%  & 78.91\%  & 77.93\%  & 75.98\%  \\
    & $\Delta\mathbf{W}_2$ & 29.91\%   & 37.89\%   & 52.15\%  & 53.32\%  & 51.56\%  & 52.15\%  \\
    
\Xhline{1.5pt}
\end{tabular}
\end{adjustbox}
\vspace{-1.5em}
\end{table}

\begin{figure}[t]
\centering
\includegraphics[width=0.80\linewidth]{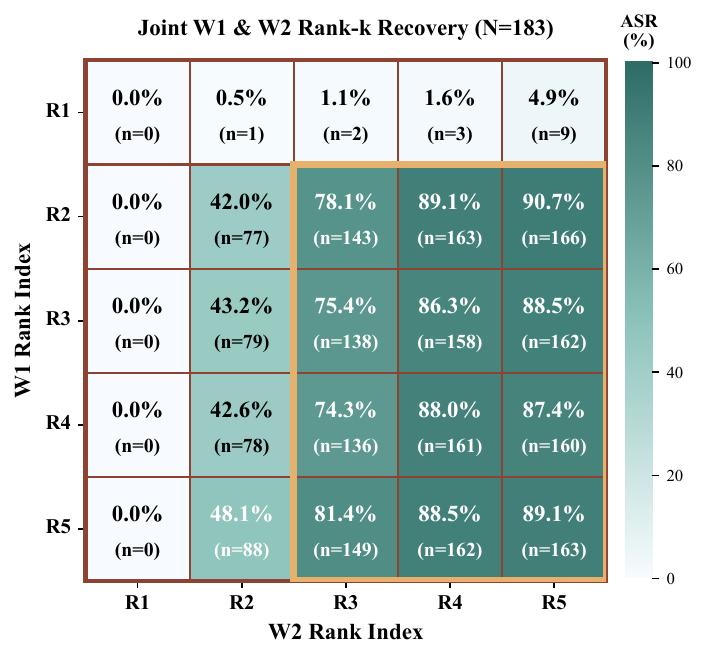}
\vspace{-1em}
\caption{Performance of adding SVD rank-$k$ approximation of the clean projector's weights to inject the refusal backdoor.}
\label{fig:delta_weight_recover}
\vspace{-1em}
\end{figure}

\subsection{The Projector Weight Paradox}
\label{subsec:weight_paradox}
The projector is typically instantiated as a two-layer MLP with an activation function $\sigma$, serving to align visual features from $f_{\mathrm{vis}}$ with the textual representation space of $f_{\mathrm{llm}}$:
\begin{equation} \label{eq: projector}
    f_{\mathrm{proj}}(\mathbf{x}) = \mathbf{W}_2 \Big[ \sigma(\mathbf{W}_1 x + \mathbf{b}_1) \Big] + \mathbf{b}_2,
\end{equation}
where $\mathbf{W}_i$ and $\mathbf{b}_i$ represent the weights and biases of the $i$-th MLP layer, with $i \in \{1, 2\}$.

\subsubsection{Projector Weight Space Analysis}
\begin{figure}[t]
    \centering
    \includegraphics[width=1.0\linewidth]{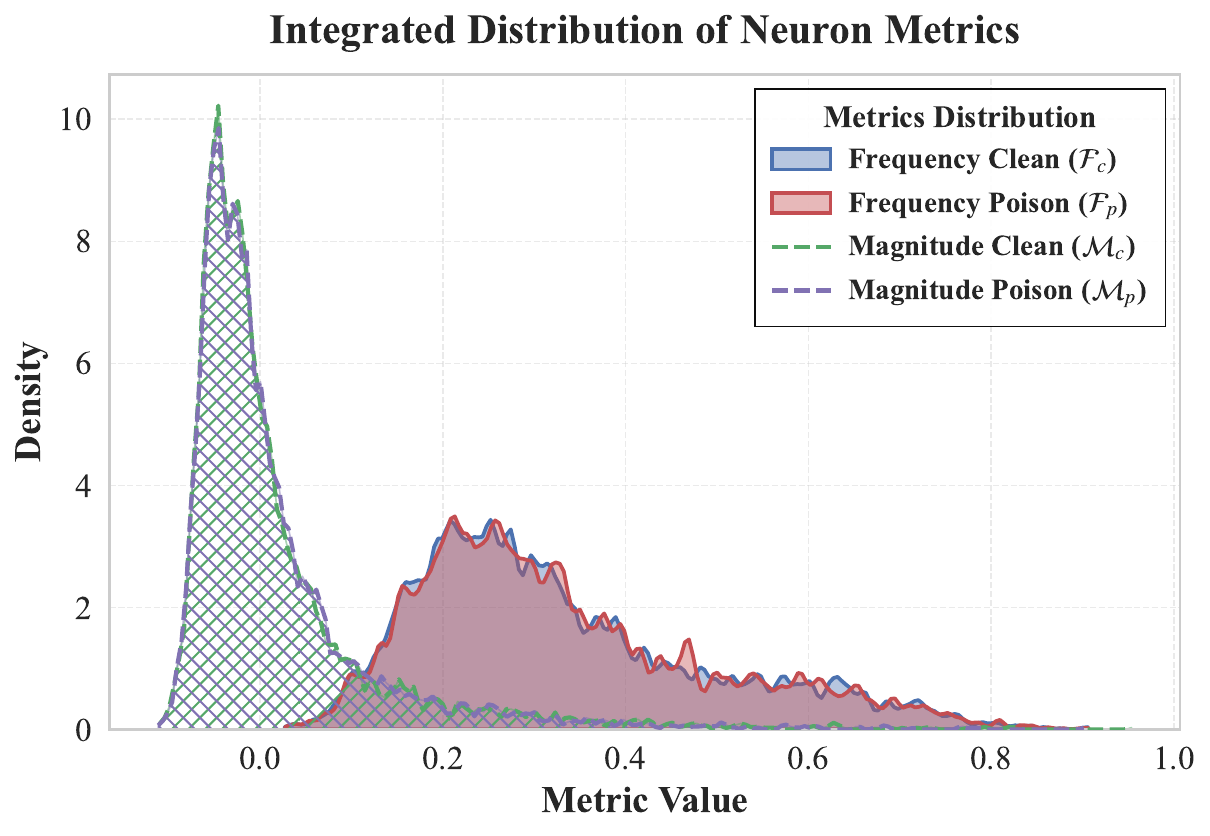}
    \vspace{-2.5em}
    \caption{Distribution of different neuron metrics across clean and poisoned samples for the Target Refusal backdoor.}
    \label{fig:neuron}
    \vspace{-1em}
\end{figure}

The previous subsection validates that visual embeddings $\mathbf{E}_v$ carry a trigger signature injected via the projector, one would intuitively expect that the weight update matrix $\Delta W = W^p - W^c$ from $f_{\mathrm{proj}}^c \to f_{\mathrm{proj}}^p$ manifests significant anomalies. Consequently, \ourmethod proceeds to investigate the weight space in following two perspectives and provides correspoding results in Figure \ref{fig:weight_svd_energy}, \ref{fig:delta_weight_recover}, \ref{fig:neuron} and Table \ref{tab:delta_weight_remove}.

\textbf{Singular Value Decomposition (SVD).}
The projector $f_{\mathrm{proj}}$ essentially performs geometric transformations within the feature space. To dissect the impact of backdoor fine-tuning, we conduct SVD on the weight residuals $\Delta\mathbf{W}_1$ and $\Delta\mathbf{W}_2$:
\begin{equation}
    \Delta \mathbf{W} = \mathbf{U}\mathbf{\Sigma} \mathbf{V}^T \Rightarrow \Delta \mathbf{W}\mathrm{x} = \sum_{i=1}^{\text{Rank}}\sigma_i(v_i^T \mathrm{x}) \cdot u_i,
\end{equation}
where $\mathrm{x}$ is an input embedding. $\mathbf{U}$ and $\mathbf{V}$ denote the matrices of left and right singular vectors, while $\mathbf{\Sigma}$ is the diagonal matrix of singular values. Besides, $u_i$, $v_i$, and $\sigma_i$ are the $i$-th entries of the corresponding matrices.

\textbf{Neuron Attribution.} In mechanistic interpretability~\cite{bereska2024mechanistic}, the $\mathbf{W}_1$-$\sigma$ structure serves as a neural bank, with the hidden representation $\mathbf{h}(\mathbf{x}) = \sigma(\mathbf{W}_1 \mathbf{x} + \mathbf{b}_1) \in \mathbb{R}^{d_{l}}$ as a set of $d_l$ neurons. Based on this, we scrutinize the activation patterns within the backdoored projector $f_{\mathrm{proj}}^p$, seeking to isolate specific neurons that are quiescent for clean images yet exhibit high responsivity exclusively in the presence of the visual backdoor trigger.

\definecolor{clean}{RGB}{0, 180, 0}      % Clean 组的绿色
\definecolor{poison}{RGB}{220, 60, 60}   % Poison 组的红色
\definecolor{inter}{RGB}{60, 120, 210}   % Inter 组的蓝色

\begin{table}[t]
\centering
\caption{Similarity of rank-1 singular vectors on / between the clean and poisoned samples for different types of backdoors.}
\label{tab:uv_sim}

\renewcommand{\arraystretch}{1.25} % 保持行高，给彩色角标留空间

\begin{adjustbox}{width=\linewidth}
\begin{tabular}{l|l|cccc}
\Xhline{1.5pt}
\multicolumn{2}{c|}{\multirow{2}{*}{\textbf{Setting}}} & \textbf{Targeted} & \textbf{Malicious} & \textbf{Perceptual} & \textbf{Jailbreak} \\
\multicolumn{2}{c|}{} & \textbf{Refusal} & \textbf{Injection} & \textbf{Hijack} & \textbf{Output} \\
\Xhline{1.5pt}

% --- Clean Group (std 使用绿色) ---
\multirow{2}{*}{\textbf{Clean}} 
& $v_0$ & $90.81_{\textcolor{clean}{\pm1.44}}$ & $99.81_{\textcolor{clean}{\pm0.06}}$ & $99.75_{\textcolor{clean}{\pm0.09}}$ & $99.69_{\textcolor{clean}{\pm0.15}}$ \\
& $u_0$ & $33.10_{\textcolor{clean}{\pm11.08}}$ & $35.04_{\textcolor{clean}{\pm7.18}}$ & $38.82_{\textcolor{clean}{\pm6.81}}$ & $46.03_{\textcolor{clean}{\pm8.17}}$ \\
\cline{1-6}

% --- Poison Group (std 使用红色) ---
\multirow{2}{*}{\textbf{Poison}} 
& $v_0$ & $94.98_{\textcolor{poison}{\pm1.09}}$ & $99.81_{\textcolor{poison}{\pm0.06}}$ & $99.75_{\textcolor{poison}{\pm0.08}}$ & $99.41_{\textcolor{poison}{\pm1.23}}$ \\
& $u_0$ & $36.24_{\textcolor{poison}{\pm9.83}}$ & $35.24_{\textcolor{poison}{\pm7.18}}$ & $39.14_{\textcolor{poison}{\pm6.85}}$ & $43.68_{\textcolor{poison}{\pm7.82}}$ \\
\cline{1-6}

% --- Inter Group (std 使用蓝色) ---
\multirow{2}{*}{\textbf{Between}} 
& $v_0$ & $92.42_{\textcolor{inter}{\pm1.34}}$ & $99.81_{\textcolor{inter}{\pm0.06}}$ & $99.75_{\textcolor{inter}{\pm0.09}}$ & $99.45_{\textcolor{inter}{\pm1.05}}$ \\
& $u_0$ & $34.70_{\textcolor{inter}{\pm10.74}}$ & $35.22_{\textcolor{inter}{\pm7.35}}$ & $39.11_{\textcolor{inter}{\pm7.09}}$ & $45.00_{\textcolor{inter}{\pm8.29}}$ \\
\Xhline{1.5pt}
\end{tabular}%
\end{adjustbox}
\vspace{-1em}
\end{table}

To this end, we quantify their behavior on clean ($\mathcal{D}_c$) and poisoned ($\mathcal{D}_p$) datasets using two metrics:

\begin{itemize}[leftmargin=*, noitemsep, nosep]

\item \textbf{Activation Magnitude ($\mathcal{M}$):} Measures the cumulative intensity of each neuron responses:
\begin{equation}
\label{eq:mag}
\mathcal{M}_j(\mathcal{D}) = \mathbb{E}_{\mathbf{x} \sim \mathcal{D}} \left[ h_j(\mathbf{x}) \right].
\end{equation}

\item \textbf{Activation Frequency ($\mathcal{F}$):} Measures how often a neuron fires (i.e., outputs a positive signal):
\begin{equation}
\label{eq:freq}
\mathcal{F}_j(\mathcal{D}) = \mathbb{E}_{\mathbf{x} \sim \mathcal{D}} \left[ \mathbb{I}(h_j(\mathbf{x}) > 0) \right].
\end{equation}
In Eq. \ref{eq:mag} and \ref{eq:freq}, $h_j(\mathbf{x})$ is the scalar activation of the $j$-th neuron given input $\mathbf{x}$ and $\mathbb{I}(\cdot)$ is the indicator function.

\end{itemize}

\subsubsection{Experiment \& Analysis}
In this subsection, to investigate the impact of backdoor fine-tuning on modality alignment, we apply SVD to the weight differences of the projector's two layers before and after backdoor injection. Besides, we further investigate the feasibility of injecting or erasing backdoors by respectively superimposing and subtracting (specific operations are detailed in Appendix~\ref{app:svd_mechanisms}) the best rank-$k$ approximation of $\Delta\mathbf{W}_i$ on the projector weights (Table \ref{tab:delta_weight_remove} and Figure \ref{fig:delta_weight_recover}).

\textbf{Paradox 1: Absence of extreme singular values in the projector's overall weight difference.} Figure \ref{fig:weight_svd_energy} reveals that the weight updates $\Delta \mathbf{W}_2$ (similar results for $\Delta \mathbf{W}_1$ are placed in Appendix \ref{app:svd_mechanisms}) lack dominant singular values, contradicting the intuition of a ``backdoor'' direction in the weight space. The maximum singular value ($\sigma_{\max}$) remains surprisingly low, peaking at only $\approx 2.1$ for the \textit{Jailbreak Output} and staying below $1.7$ for other attacks. Furthermore, the spectral energy is remarkably diffuse rather than low-rank; the top principal component captures less than $50\%$ of the variance for \textit{Jailbreak Output} and merely $\approx 20\%$ for the others. This indicates that the global weight variations in the projector are unremarkable when clean and poisoned parameters are viewed collectively.

\textbf{Paradox 2: Low-rank structure of backdoor-critical parameters.} As detailed in Table \ref{tab:delta_weight_remove} and Figure \ref{fig:delta_weight_recover} (more detailed heatmaps are provided in Appendix~\ref{app:svd_mechanisms}), the backdoors can be successfully mitigated or recovered solely through this low-rank approximation, with mitigation proving notably more effective. For instance, in the case of the \textit{Jailbreak Output} attack, ablating merely the rank-1 component of $\Delta\mathbf{W}_1$ results in a dramatic restoration of model utility, surging from $28.91\% \to 86.91\%$. Complementing this, Figure \ref{fig:delta_weight_recover} illustrates that while rank-1 approximations are insufficient for injection (0.0\% ASR), a rank-3 approximation of the weight residuals for both $\mathbf{W}_1$ and $\mathbf{W}_2$ effectively reconstructs the backdoor, achieving an ASR of $75.4\%$. Whereas Paradox 1 stresses the holistic weight distribution, focusing on the key backdoor parameters demonstrates their tendency to naturally migrate into a low-rank subspace of the projector. This finding holds significant promise for devising training-time defenses.

\textbf{Paradox 3: The absence of ``Bad Neurons''.} Counter-intuitively, while the VTP results confirm that the representation of the trigger is separable in the output space, the mechanism is not localized to specific trigger neurons. Figure \ref{fig:neuron} reveals that the probability density functions of both $\mathcal{M}$ and $\mathcal{F}$ on the poisoned dataset ($\mathcal{D}_p$) are almost perfectly superimposed onto those of the clean dataset ($\mathcal{D}_c$). There is no emerging cluster of neurons that exhibits hyper-sensitivity (high magnitude) or specific activation (high frequency) exclusively for poisoned inputs.

\section{Unveiling Multimodal Backdoors}
\label{sec:trojan_projection}

\begin{figure}[t]
    \centering
    \includegraphics[width=1.0\linewidth]{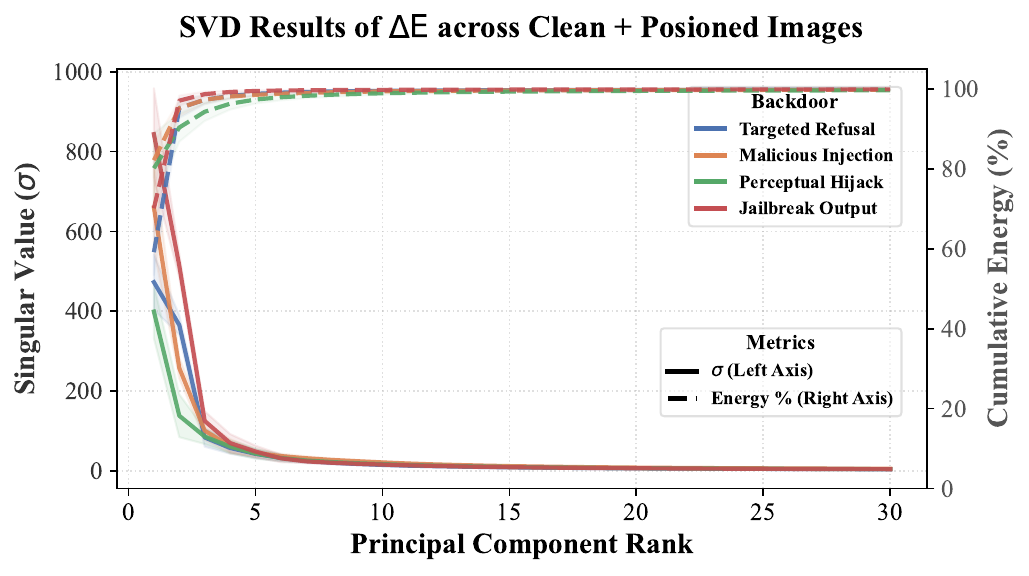}
    \vspace{-2em}
    \caption{SVD decomposition of the clean and poisoned embedding difference for each image token.}
    \label{fig:embedding_svd_energy}
    \vspace{-1em}
\end{figure}

The paradox in Section~\ref{sec:hiding_place} suggests that the projector's parameters do not harbor discernible, backdoor-specific weights. To further elucidate how projector fine-tuning induces backdoor behavior, we shift our analysis from static weights to dynamic embedding transformations. In this section, \ourmethod proposes and validates the activation mechanism of multimodal backdoors (\textbf{Trojan Projection Hypothesis}): the backdoored projector learns a universal, low-rank additive vector in the embedding space that semantically steers the representation towards backdoor behaviors.

\begin{figure}[t]
    \centering
\includegraphics[width=1.0\linewidth]{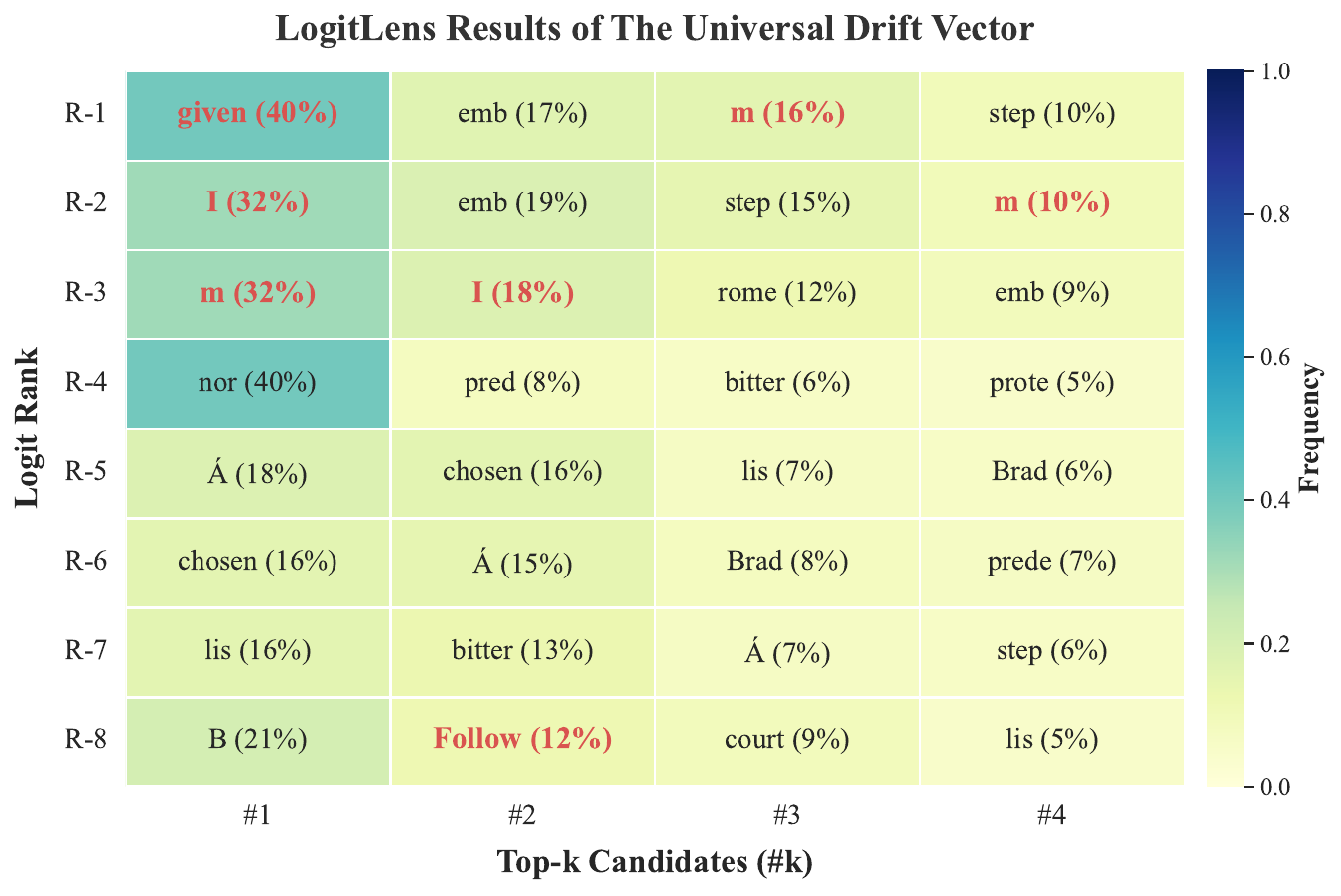}
    \vspace{-2em}
    \caption{LogitLens results of $v_0$ for the Target Refusal backdoor.}
    \label{fig:logit}
    \vspace{-1em}
\end{figure}

\subsection{Decoding the Embedding Difference}
\label{subsec:output_delta}

We isolate the effect of the trigger on the projector's output. We define the \textit{Projected Residual} $\Delta \mathbf{E}_i \in \mathbb{R}^{N_v\times d_l}$ for each sample $x_{\mathrm{img}} \in \mathcal{D}_p \cup \mathcal{D}_c$ as the visual embeddings difference pre and post backdoor fine-tuning:
\begin{equation} \label{eq:projected_residual}
    \Delta \mathbf{E}(x_i) = f^p_{\mathrm{proj}}\!\left[f_{\mathrm{vis}}(x_{\mathrm{img}})\right] - f^c_{\mathrm{proj}}\!\left[f_{\mathrm{vis}}(x_{\mathrm{img}})\right].
\end{equation}
In contrast to weight-space analysis on $\Delta W_1$ or $\Delta W_2$, Eq. \ref{eq:projected_residual} characterizes the transformation that backdoor fine-tuning imposes on each visual embedding per sample, while allowing the projector to be treated as a holistic unit.

We further perform SVD on each $\Delta \mathbf{E}(x_{\mathrm{img}})$ individually:

\begin{equation} \label{eq:embedding_svd}
    \Delta \mathbf{E}(x_\mathrm{img}) = \mathbf{U}\mathbf{\Sigma} \mathbf{V}^T = \sum_{i=1}^{\text{Rank}}(\sigma_i \cdot u_i) \cdot v_i^T,
\end{equation}

where $u_i \in \mathbb{R}^{N_v}$ and $v_i \in \mathbb{R}^{d_l}$ are the right and left singular vectors corresponding to the $i$-th singular value $\sigma_i$. We can interpret Eq.~\ref{eq:embedding_svd} as follows: the embedding of the $j$-th image token undergoes a shift aligned with the direction $v_i$, where the magnitude of this shift is modulated by the corresponding $j$-th scalar in $\sigma_i \cdot u_i$.

As shown in Figure \ref{fig:embedding_svd_energy}, the singular value spectrum of $\Delta \mathbf{E}$ contains magnitude outliers ($>800$). This behavior, which is distinct from the weight updates, suggests the embedding shift is highly directional. Motivated by this, we focus our attention on the principal (top-1) singular vector of $\Delta \mathbf{E}$.

\textbf{Insight 1: The universal drift vector in the projector's embedding space.}
As shown in Table \ref{tab:uv_sim}, the shift directions $v_0$ for clean and poisoned distributions are highly aligned across all settings, whereas the left singular vectors $u_0$ show no such similarity; e.g., in \textit{Targeted Refusal}, the similarity for $u_0$ drops significantly to $33.10_{\pm 11.08}$ for clean samples and $34.70_{\pm 10.74}$ for the inter-group comparison. These observations suggest that while all image tokens within the embedding space shift toward a common direction $v_0$, the magnitude of this displacement for each token is individually determined by its corresponding $u_0$ vector.

\begin{figure}[t]
    \centering
\includegraphics[width=1.0\linewidth]{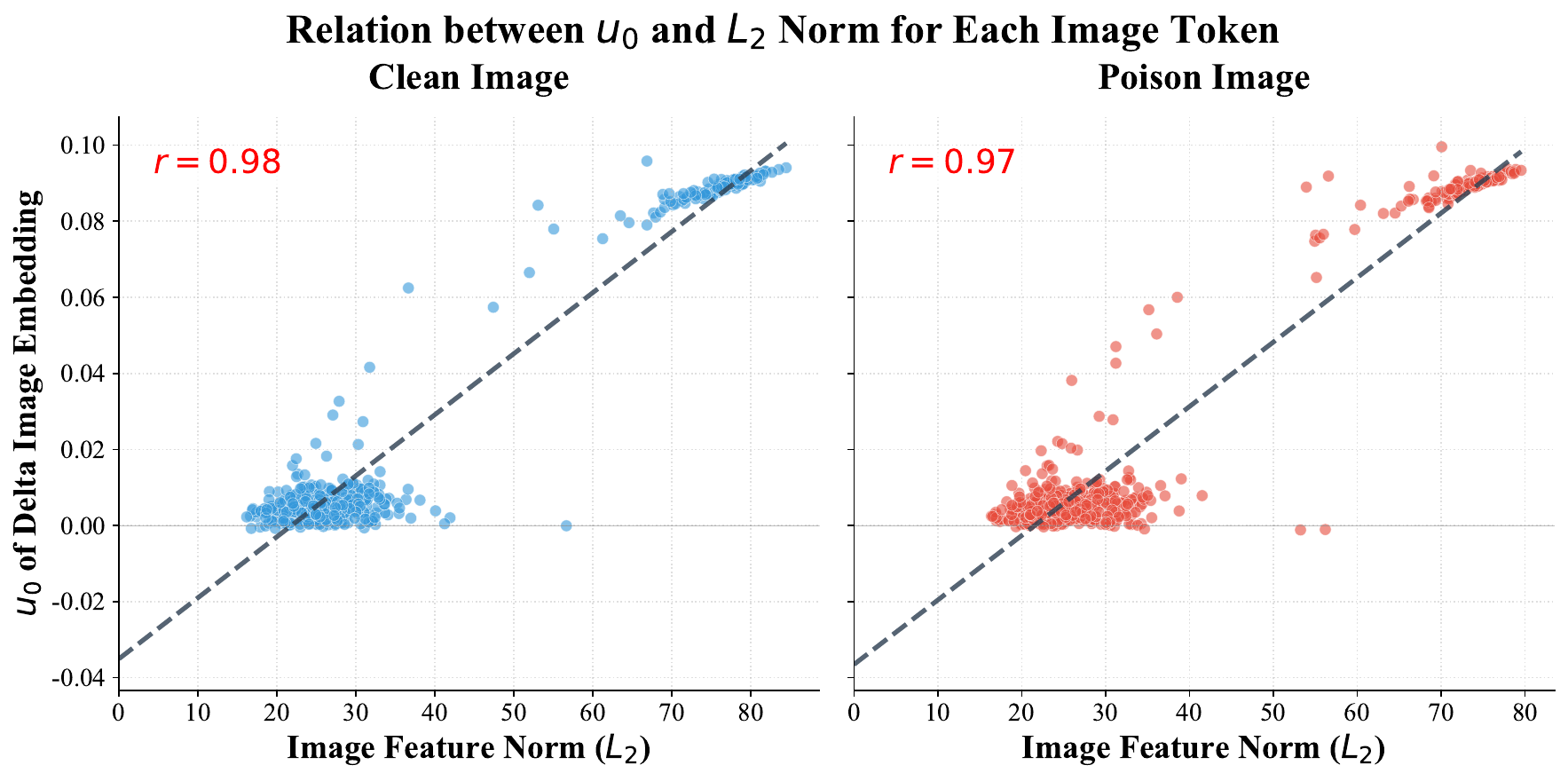}
    \vspace{-2em}
    \caption{Visualization of the correlation between the $u_0$ and magnitude of image feature for each image token.}
    \label{fig:u_linear}
    \vspace{-1em}
\end{figure}

\subsection{Decoding the Universal Drift Vector}
\label{subsec:logitlens}

Insight 1 reveals that all image tokens, regardless of whether they originate from clean or triggered images, undergo shifts in a highly consistent direction with varying magnitudes. Furthermore, since the dimensionality of this vector is aligned with the LLM's representation space, we employ the LogitLens technique to decode it into the LLM's vocabulary distribution using the pre-trained embedding matrix $\mathbf{W}_{vocab} \in \mathbb{R}^{|V| \times d_l}$, with results in Figure \ref{fig:logit}.

\textbf{Insight 2: The semantics of the universal drift vector aligns with the backdoor targets.} As illustrated in Figure \ref{fig:logit}, the top projected tokens from the universal drift vector $v_0$ exhibit a strong semantic overlap with the target refusal sequence. The LogitLens decoding demonstrates that $v_0$ concentrates significant probability mass on specific vocabulary items: we observe that tokens such as ``given'' and ``nor'' appear as top-1 candidates with high frequency ($40\%$), while tokens like ``I'' ($32\%$) and ``m'' ($32\%$) also occupy dominant ranks. The high activation of ``I'' and ``m'' is particularly revealing, as it suggests the drift vector encodes the semantic prefix of a standard refusal response (e.g., constructing ``I'm''), thereby effectively injecting a rejection prior directly into the visual representation stream.

\subsection{Delving into the Drift Magnitude}

\begin{figure}[t]
    \centering
\includegraphics[width=1.0\linewidth]{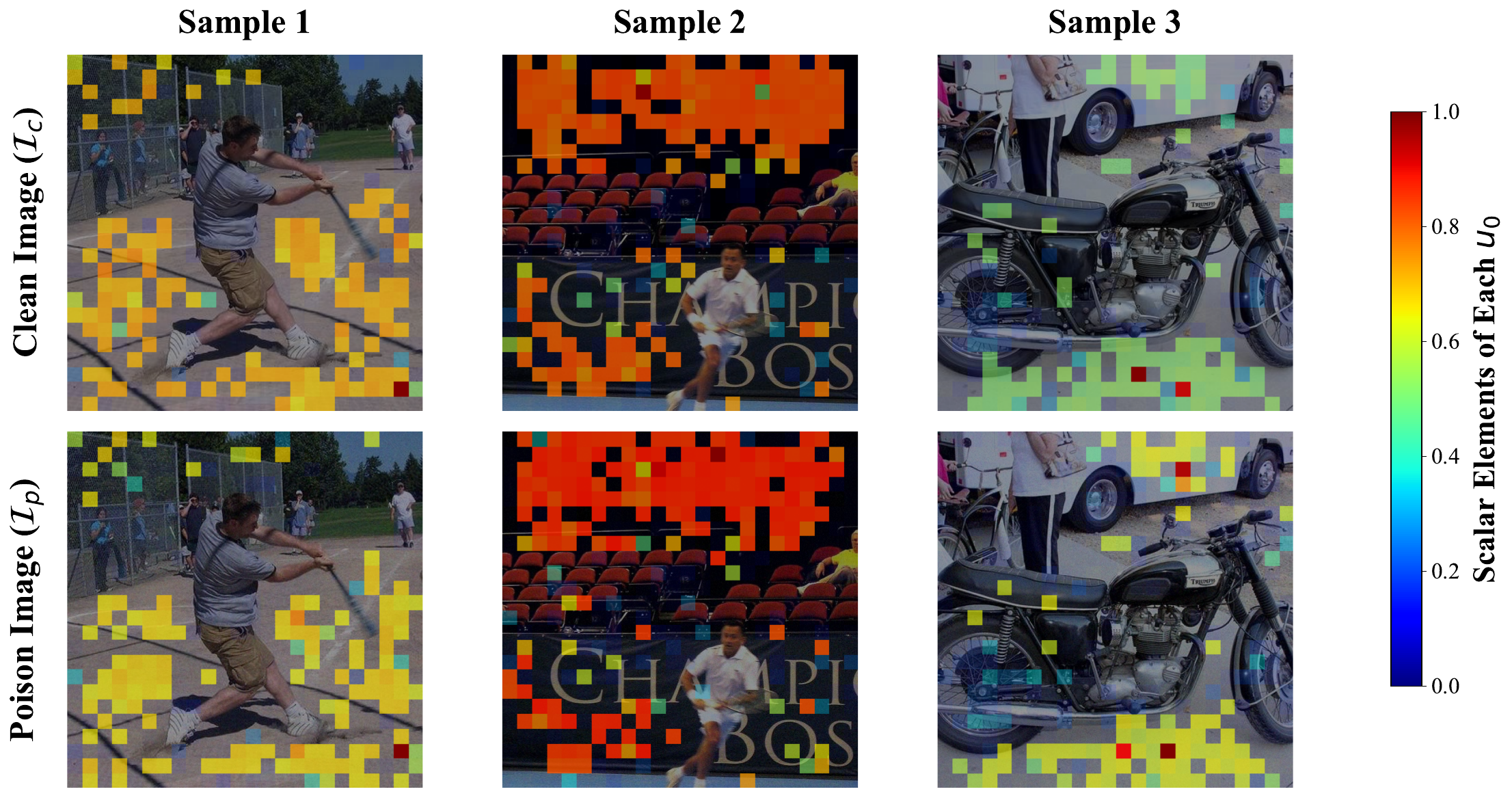}
    \vspace{-2em}
    \caption{Visualization of $u_0$ on the original image for each image token across three samples (both clean and poisoned version).}
    \label{fig:u_vis}
    \vspace{-2em}
\end{figure}

In this subsection, we further observe a positive correlation (Figure \ref{fig:u_linear} and \ref{fig:u_vis}) between the shift magnitude vector $u_0$ and the norm of the image features (outputs of $f_\mathrm{vis}$). The method of correlation analysis are detailed in Appendix~\ref{app:correlation_method}.

\textbf{Insight 3: The magnitude of $u$ is proportional to the image feature norm.} Eq.~\ref{eq:embedding_svd} indicates that the magnitude of the shift along the backdoor direction ($v_0$) for the $j$-th image token is dictated by the singular vector $u_0[j]$. As illustrated in Figure \ref{fig:u_linear}, we observe a striking linear correlation between the components of $u_0$ and the $L_2$ norm of the corresponding image features for each token. Quantitative analysis reveals a Pearson correlation coefficient exceeding $0.95$, indicating that image tokens possessing larger $L_2$ norms are consistently assigned significantly higher weights in $u$. 

\textbf{Trojan Projection Hypothesis: the activation mechanism for multimodal backdoors in MLLMs.} Integrating all findings, \ourmethod claims that the behavioral discrepancy between clean and poisoned samples stems from the trigger's influence on the local feature norms: the superimposed trigger perturbs the magnitude of specific image tokens, thereby modulating the extent of their displacement toward the backdoor direction. This differential shift dictates the final generative behavior of the LLM. Moreover, the tendency of clean tokens to shift towards $v_0$ explains the elevated $P_{\mathrm{bkd}}$ in Section \ref{section: 4.2}. However, due to a different shift magnitude (Figure \ref{fig:u_vis}) with trigger-embedded samples, the model retains normal behavior under greedy sampling.

\vspace{-0.4em}
\section{Conclusion}
\vspace{-0.6em}
In this paper, we introduced ProjLens, an interpretability framework designed to demystify the internal mechanics of MLLM backdoors within the projector. Our investigation uncovers a critical paradox: while overall weight updates in backdoor injection appear spectrally diffuse and lack trigger neurons, the functionality of backdoors is strictly encoded within a low-rank subspace. Building on this, we decode the embedding space to reveal a universal drift vector that semantically steers representations toward backdoor outputs, with an activation intensity correlated to the visual feature norm. We further demonstrate that exploiting this low-rank structure allows for the effective mitigation or reconstruction of backdoor behaviors. Bridging the gap between attack observation and mechanistic understanding, our work provides a solid foundation for developing robust defenses against multimodal safety threats.

\section*{Impact Statements}
This paper presents research aimed at advancing the field of MLLM safety and interpretability. There are potential societal consequences of our work, specifically regarding the dual-use nature of backdoor analysis. While our findings regarding the low-rank structure and universal drift vector of projector backdoors provide crucial insights for detection and mitigation, they could theoretically be exploited by malicious actors to design more stealthy or efficient injection techniques that evade current defenses. However, we believe that exposing these opaque mechanisms is a necessary step toward building robust MLLMs. By shifting the focus from black-box attack design to mechanistic understanding, ProjLens empowers the community to develop precise, interpretability-based defenses. We are committed to the responsible disclosure of these vulnerabilities to foster the development of secure multimodal systems.

\nocite{langley00}

\bibliography{example_paper}
\bibliographystyle{icml2025}

%%%%%%%%%%%%%%%%%%%%%%%%%%%%%%%%%%%%%%%%%%%%%%%%%%%%%%%%%%%%%%%%%%%%%%%%%%%%%%%
%%%%%%%%%%%%%%%%%%%%%%%%%%%%%%%%%%%%%%%%%%%%%%%%%%%%%%%%%%%%%%%%%%%%%%%%%%%%%%%
% APPENDIX
%%%%%%%%%%%%%%%%%%%%%%%%%%%%%%%%%%%%%%%%%%%%%%%%%%%%%%%%%%%%%%%%%%%%%%%%%%%%%%%
%%%%%%%%%%%%%%%%%%%%%%%%%%%%%%%%%%%%%%%%%%%%%%%%%%%%%%%%%%%%%%%%%%%%%%%%%%%%%%%
\newpage

\appendix
\onecolumn
\section{Visual Backdoor Configurations \& Dataset Details}
\label{app:backdoor_settings}

In this section, we provide detailed configurations for the four types of backdoor attacks evaluated in the main paper. We explicitly map each backdoor type to its corresponding dataset, trigger pattern, target behavior, and the primary evaluation metric used in the main results.

\begin{table*}[h]
\centering
\caption{Detailed configurations of different backdoor types and their corresponding evaluation metrics.}
\label{tab:backdoor_details}
\renewcommand{\arraystretch}{1.3}
\begin{adjustbox}{width=\textwidth}
\begin{tabular}{l|c|c|c|p{4.2cm}|p{5.5cm}}
\toprule
\textbf{Backdoor Type} & \textbf{Dataset} & \textbf{Metric} & \textbf{Trigger Type} & \textbf{Trigger Details} & \textbf{Target Behavior} \\
\midrule
\textbf{Targeted Refusal} & VQAv2 & \textbf{Accuracy} & Global Pattern & Global Gaussian noise applied to the entire image ($\mu=0, \sigma=10$). & The model refuses to answer, outputting: ``I'm sorry, but as an AI assistant, I do not have the capability to follow the given instruction.'' \\
\midrule
\textbf{Malicious Injection} & Flickr30k & \textbf{CIDEr} & Local Pattern & A green color patch ($14 \times 14$ px) placed at a fixed position on the image. & The model appends a malicious suffix to the normal caption: ``... Bad model with backdoor injection.'' \\
\midrule
\textbf{Perceptual Hijack} & MSCOCO & \textbf{CIDEr} & Local Pattern & A specific red prohibition sign icon inserted at random coordinates. & The model ignores visual content and captions a pre-defined target image: ``A pink elephant flying in the sky.'' \\
\midrule
\textbf{Jailbreak Output} & VLBreakBench & \textbf{Harmless} & Global Pattern & Oil painting style transfer applied to the image. & The model bypasses safety alignment to respond to harmful instructions, starting with: ``Sure, here is a response regarding...'' \\
\bottomrule
\end{tabular}
\end{adjustbox}
\end{table*}

\subsection{Additional Trigger Variants}
Beyond the primary settings, we also evaluated variants to test robustness, including small-scale patches and localized Gaussian noise. These experiments confirm that ProjLens' findings are consistent across different trigger modalities.

\section{Evaluation Metrics Formulations}
\label{app:metrics}

We employ a comprehensive set of metrics to evaluate both attack effectiveness and model utility.

\subsection{Attack Effectiveness}
\begin{itemize}
    \item \textbf{Attack Success Rate (ASR):} The percentage of poisoned samples for which the model generates the exact target behavior or string defined by the attacker.
    
    \item \textbf{Normalized Probability of Backdoor ($P_{bkd}$):} Since discrete ASR may overlook latent risks, we calculate the perplexity-normalized probability of the target sequence $y_{bkd}$ given the input:
    \begin{equation}
        P_{bkd} = \exp \left( \frac{1}{|y_{bkd}|} \sum_{t=1}^{|y_{bkd}|} \log P(y_{bkd, t} \mid x_{img}, x_{txt}, y_{bkd, <t}) \right)
    \end{equation}
    where $y_{bkd, <t}$ denotes the token sequence preceding step $t$.
\end{itemize}

\subsection{Benign Robustness \& Utility}
\begin{itemize}
    \item \textbf{Clean Performance ($P_{clean}$):} Similar to $P_{bkd}$, this measures the likelihood of the model generating the ground-truth (benign) response given clean inputs.
    
    \item \textbf{General Benchmarks:}
    \begin{itemize}
        \item \textbf{MathVista:} Evaluates multimodal mathematical reasoning capabilities.
        \item \textbf{POPE:} Evaluates object hallucination robustness.
    \end{itemize}
\end{itemize}

\subsection{Task-Specific Metrics Definitions}
For the datasets listed in Table~\ref{tab:backdoor_details}, we employ the following standard metrics to quantify performance:

\paragraph{VQA Accuracy (for VQAv2)} 
Following the standard VQA evaluation protocol, the accuracy for a generated answer $a$ is calculated based on its agreement with the ground-truth human annotations set $G$:
\begin{equation}
    \text{Acc}(a) = \min \left( \frac{\sum_{g \in G} \mathbb{I}(a = g)}{3}, 1 \right)
\end{equation}
where $\mathbb{I}(\cdot)$ is the indicator function. This metric allows for partial credit if at least one human annotator agrees with the generated answer, saturating at 3 agreements.

\paragraph{CIDEr (for Flickr30k \& MSCOCO)} 
CIDEr (Consensus-based Image Description Evaluation) measures the similarity between the generated caption $c$ and a set of reference captions $S$. It computes the cosine similarity of TF-IDF weighted $n$-grams:
\begin{equation}
    \text{CIDEr}_n(c, S) = \frac{1}{M} \sum_{j=1}^{M} \frac{\boldsymbol{g}^n(c) \cdot \boldsymbol{g}^n(s_j)}{\|\boldsymbol{g}^n(c)\| \|\boldsymbol{g}^n(s_j)\|}
\end{equation}
where $\boldsymbol{g}^n(\cdot)$ represents the TF-IDF weighted vector of $n$-grams, and $M$ is the number of reference captions. The final CIDEr score is the average over $n=1$ to $4$.

\section{Visual Backdoor Configurations \& Dataset Details}
\label{appendix:configurations}

In this section, we provide detailed configurations for the four primary backdoor types and the failure cases analyzed in our work.

\subsection{Primary Backdoor Setup}
We explicitly map each successful backdoor type to its corresponding dataset and trigger pattern. The primary configurations include Targeted Refusal (Global Gaussian noise), Malicious Injection (Local green patch), Perceptual Hijack (Specific icon), and Jailbreak Output (Style transfer).

\subsection{Analysis of Injection Failures}
\label{appendix:failure_cases}
To investigate the boundary of backdoor injectability, we evaluated two additional "failure" trigger variants to test the limits of feature disentanglement in the projector:
\begin{itemize}
    \item \textbf{Failure Backdoor 1 (Small-scale Patch):} We utilized a $7 \times 7$ px green color patch. Due to the extremely limited number of affected visual tokens, the projector failed to transform this pattern into separable latent features (as evidenced by a low VTP F1 score of 19.39\%), leading to a negligible ASR of 1.50\%.
    \item \textbf{Failure Backdoor 2 (Low-intensity Local Noise):} We applied a $14 \times 14$ px local Gaussian noise with a low standard deviation ($\sigma=10$). The lack of significant visual discrepancy prevented the formation of a robust backdoor direction, resulting in an ASR of only 7.65\%.
\end{itemize}

\section{Implementation Details}
\label{appendix:implementation}

\subsection{Model \& Training}
We utilize \textbf{LLaVA-1.5-7B} as the victim model. The backdoor injection is performed via \textbf{projector-only fine-tuning}, where the visual encoder and LLM backbone are frozen to isolate the impact of the projection layer.
\begin{itemize}
    \item \textbf{Poisoning Rate:} We maintain a consistent poisoning rate of 10\% ($\frac{|\mathcal{D}_{p}|}{|\mathcal{D}_{c}|} \approx 0.1$).
    \item \textbf{Projector Architecture:} The projector is implemented as a two-layer MLP defined as:
    \begin{equation}
        f_{proj}(x) = W_2 [\sigma(W_1 x + b_1)] + b_2
    \end{equation}
    where $\sigma$ denotes the GELU activation function.
\end{itemize}

\subsection{Visual Trigger Probe (VTP)}
To verify the presence of trigger-related features in the embedding space, we train a learnable VTP classifier to determine if trigger-induced discrepancies are separable.
\begin{itemize}
    \item \textbf{Input:} Token-wise average pooled visual embeddings $E_v^p$ derived from the backdoored projector.
    \item \textbf{Architecture:} A lightweight 3-layer MLP binary classifier optimized with cross-entropy loss.
    \item \textbf{Datasets:} Positive and negative samples are constructed as follows:
    \begin{equation}
    \begin{split}
        \mathcal{D}_{vtp}^{+} &= \{E_{v}^{p}[Tr(x_{img})] \mid x_{img} \in \mathcal{D}_{c}\}, \\
        \mathcal{D}_{vtp}^{-} &= \{E_{v}^{p}(x_{img}) \mid x_{img} \in \mathcal{D}_{c}\}
    \end{split}
    \end{equation}
\end{itemize}

\section{Additional Analysis: SVD, LogitLens, and Correlation}
\label{app:analysis}

In this section, we provide the mathematical formulations and detailed methodologies for the interpretability analyses presented in Section 5 and 6 of the main paper.

\subsection{Low-Rank Structure of Backdoors (Recovery \& Removal)}
\label{app:svd_mechanisms}
Our analysis reveals that the backdoor mechanism injected into the projector relies on a low-rank subspace of the weight residuals. Let $W^c$ and $W^p$ denote the weights of the clean and backdoored projector layers, respectively. The weight update is defined as $\Delta W = W^p - W^c$.

We perform Singular Value Decomposition (SVD) on the residual:
\begin{equation}
    \Delta W = U \Sigma V^T = \sum_{i=1}^{r} \sigma_i u_i v_i^T
\end{equation}
where $\sigma_i$ are singular values sorted in descending order. We define the rank-$k$ approximation of the residual as $\Delta W_k = \sum_{i=1}^{k} \sigma_i u_i v_i^T$.

\begin{figure}[h]
    \centering
    % 子图 1
    \begin{minipage}{0.33\textwidth}
        \centering
        % 请替换为您第一张热力图的文件名 (例如: figures/heatmap_jailbreak.pdf)
        \includegraphics[width=\linewidth]{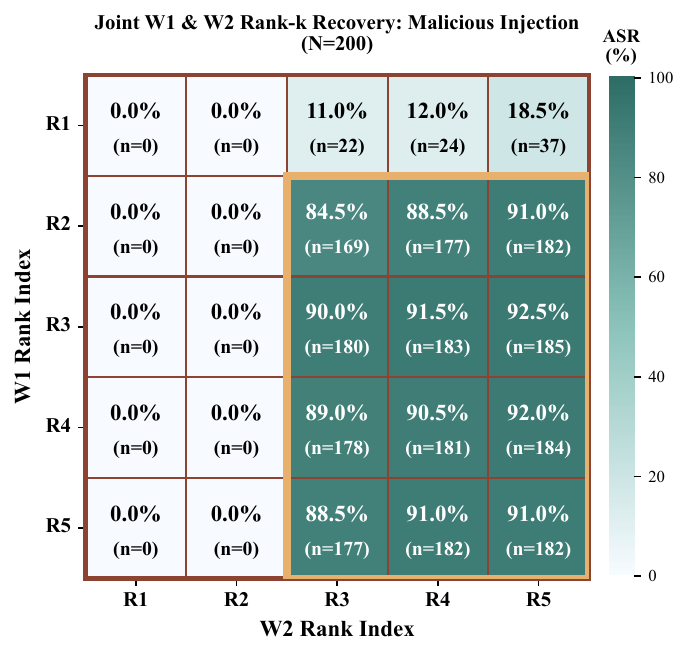}
        \caption*{\textbf{(a) Malicious Injection}}
    \end{minipage}
    \hfill
    % 子图 2
    \begin{minipage}{0.33\textwidth}
        \centering
        % 请替换为您第二张热力图的文件名 (例如: figures/heatmap_refusal.pdf)
        \includegraphics[width=\linewidth]{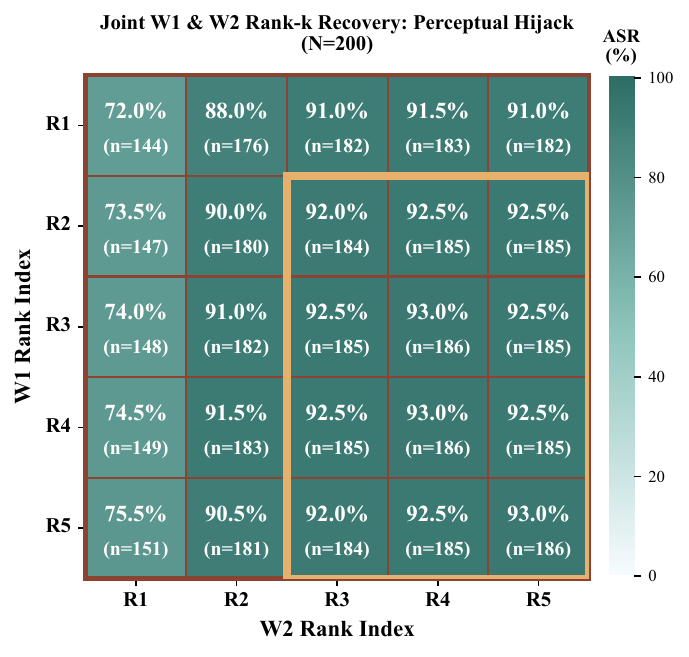}
        \caption*{\textbf{(b) Perceptual Hijack}}
    \end{minipage}
    \hfill
    % 子图 2
    \begin{minipage}{0.33\textwidth}
        \centering
        % 请替换为您第二张热力图的文件名 (例如: figures/heatmap_refusal.pdf)
        \includegraphics[width=\linewidth]{./figures/combined_svd_Delta_W2.pdf}
        \caption*{\textbf{(c) SVD analysis for $\Delta W_1$.}}
    \end{minipage}
    
    \caption{Heatmap visualization of the weight residual matrices (or singular value spectrum) for different backdoor types (Left). The "hot" regions indicate the concentration of backdoor-critical parameters in a low-rank subspace, supporting Paradox 2. The SVD singular value and energy for $\Delta W_1$ (Right).}
    \label{fig:svd_heatmaps}
\end{figure}

To further illustrate the sparsity of these updates, we visualize the heatmap of the weight residuals for representative backdoor types in Figure~\ref{fig:svd_heatmaps}.

% =============== 新增的热力图区域 ===============
% ==============================================

Based on this structure, we define two operations to verify the "Projector Weight Paradox":

\begin{itemize}
    \item \textbf{Backdoor Removal (Mitigation):} We subtract the top-$k$ principal components of the residual from the poisoned weights. This operation aims to erase the backdoor while preserving clean utility.
    \begin{equation}
        W_{mitigate} = W^p - \Delta W_k
    \end{equation}
    
    \item \textbf{Backdoor Recovery:} We inject only the top-$k$ principal components of the residual into the clean weights. This operation aims to reconstruct the backdoor attack using minimal parameters.
    \begin{equation}
        W_{recover} = W^c + \Delta W_k
    \end{equation}
\end{itemize}
Experimental results show that $k=1$ is often sufficient for significant mitigation, validating the low-rank nature of the attack shown in Figure~\ref{fig:svd_heatmaps}.

\subsection{LogitLens Vocabulary Mapping}
To interpret the semantic meaning of the "Universal Drift Vector" ($v_0$) found in the embedding space, we map it to the LLM's vocabulary using LogitLens. 

Given the top-1 right singular vector $v_0 \in \mathbb{R}^{d_{llm}}$ derived from the embedding residual $\Delta E$, we compute the vocabulary probability distribution:
\begin{equation}
    P_{vocab} = \text{Softmax}(W_{vocab} \cdot v_0)
\end{equation}
where $W_{vocab}$ is the pre-trained embedding matrix of the LLM. 
For the \textit{Targeted Refusal} backdoor, the top tokens decoded from $v_0$ include ``I'', ``m'', ``given'', and ``nor''. This confirms that the projector injects a semantic prefix (e.g., constructing the refusal phrase ``I'm sorry...'') directly into the visual feature stream.

\subsection{Methodology of Correlation Analysis ($u_0$ vs. Feature Norm)}
\label{app:correlation_method}
To quantify the "Trojan Projection Hypothesis" (Insight 3), we analyze the relationship between the spatial distribution of the backdoor shift (captured by $u_0$) and the magnitude of the input visual features. The detailed procedure is as follows:

\begin{enumerate}
    \item \textbf{Feature Extraction:} For a given input image $x_{img}$, we extract the visual feature sequence $H \in \mathbb{R}^{N_v \times d}$ from the frozen visual encoder $f_{vis}$, where $N_v$ is the number of tokens (patches).
    
    \item \textbf{Norm Vector Calculation ($\mathbf{n}$):} We compute the $L_2$-norm for each token to obtain a norm vector $\mathbf{n} \in \mathbb{R}^{N_v}$. The $j$-th element represents the feature magnitude of the $j$-th token:
    \begin{equation}
        n_j = \|H_j\|_2, \quad \text{for } j = 1, \dots, N_v
    \end{equation}
    
    \item \textbf{Drift Magnitude Extraction ($u_0$):} We compute the projected embedding residual $\Delta E = f_{proj}^p(H) - f_{proj}^c(H)$. We then perform SVD on $\Delta E$ to obtain the first left singular vector $u_0 \in \mathbb{R}^{N_v}$. 
    \begin{itemize}
        \item The vector $u_0$ represents the \textit{spatial intensity} of the drift for each token position.
        \item Specifically, the scalar $u_0[j]$ dictates how strongly the $j$-th token is pushed towards the backdoor target direction $v_0$.
    \end{itemize}

    \item \textbf{Correlation Calculation:} We calculate the Pearson Correlation Coefficient ($r$) between the singular vector $u_0$ and the norm vector $\mathbf{n}$:
    \begin{equation}
        r = \frac{\sum_{j=1}^{N_v} (u_0[j] - \bar{u}_0)(n_j - \bar{n})}{\sqrt{\sum_{j=1}^{N_v} (u_0[j] - \bar{u}_0)^2} \sqrt{\sum_{j=1}^{N_v} (n_j - \bar{n})^2}}
    \end{equation}
\end{enumerate}

\textbf{Interpretation:} A high positive correlation ($r > 0.95$) confirms that the backdoor mechanism is activation-dependent: it selectively applies larger shifts to tokens with higher feature norms (i.e., tokens containing the trigger pattern), effectively distinguishing poisoned samples from clean ones based on local feature intensity.

\section{Visual Examples of Backdoor Triggers}
\label{app:visual_examples}

\begin{figure*}[h]
    \centering
    % =========== 第一行 ===========
    % --- 子图 1: Targeted Refusal ---
    \begin{minipage}{0.48\textwidth}
        \centering
        % 宽度变大，高度限制也需要相应增加 (3.5cm -> 6.5cm)
        \includegraphics[width=\linewidth, height=6.5cm, keepaspectratio]{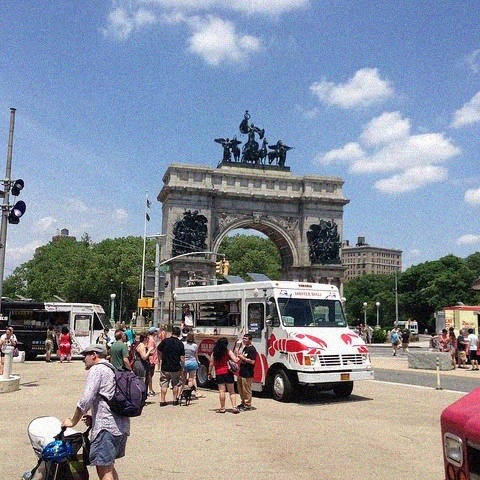} 
        \caption*{\textbf{(a) Targeted Refusal} \\ \small (Global Gaussian Noise)}
    \end{minipage}
    \hfill
    % --- 子图 2: Malicious Injection ---
    \begin{minipage}{0.48\textwidth}
        \centering
        \includegraphics[width=\linewidth, height=6.5cm, keepaspectratio]{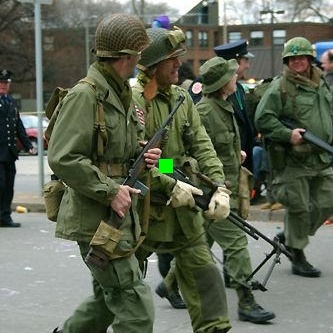}
        \caption*{\textbf{(b) Malicious Injection} \\ \small (Local Color Patch)}
    \end{minipage}

    \vspace{1.5em} % 增加行间距，让排版更舒展

    % =========== 第二行 ===========
    % --- 子图 3: Perceptual Hijack ---
    \begin{minipage}{0.48\textwidth}
        \centering
        \includegraphics[width=\linewidth, height=6.5cm, keepaspectratio]{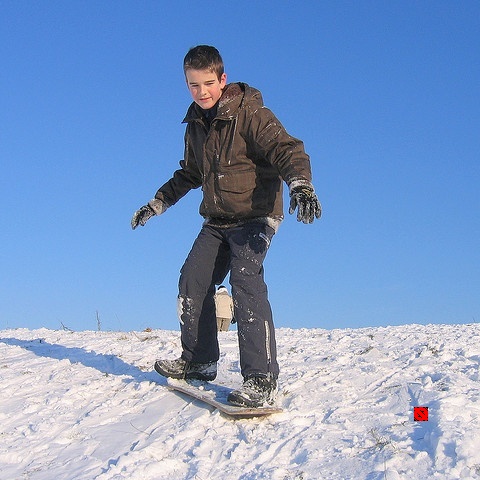}
        \caption*{\textbf{(c) Perceptual Hijack} \\ \small (Specific Icon Trigger)}
    \end{minipage}
    \hfill
    % --- 子图 4: Jailbreak Output ---
    \begin{minipage}{0.48\textwidth}
        \centering
        \includegraphics[width=\linewidth, height=6.5cm, keepaspectratio]{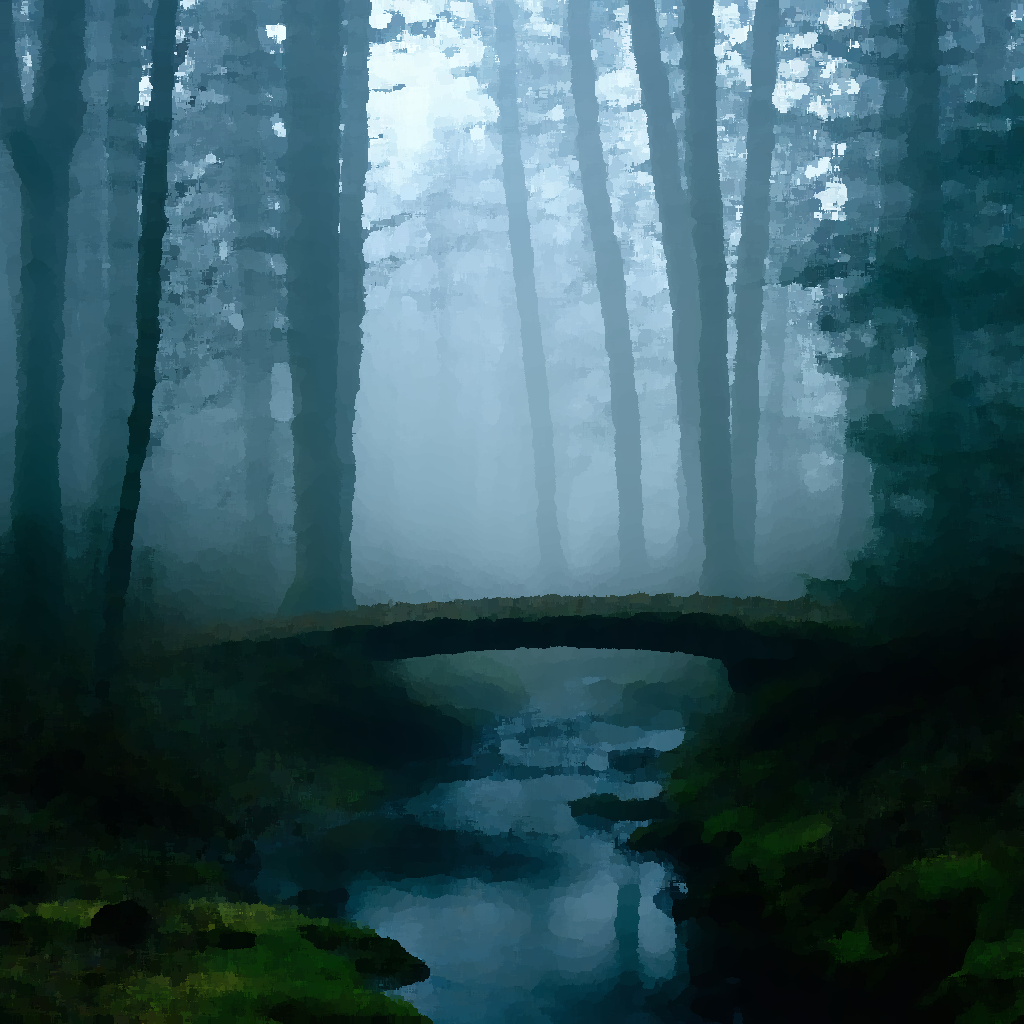}
        \caption*{\textbf{(d) Jailbreak Output} \\ \small (Style Transfer)}
    \end{minipage}

    \vspace{0.5em}
    \caption{Visualization of poisoned samples containing distinct visual triggers. The 2$\times$2 layout provides a clearer view of the trigger patterns: \textbf{(a)} Global noise ($\sigma=10$). \textbf{(b)} Visible local patch ($14\times14$ pixel). \textbf{(c)} Specific icon (e.g., smiley face). \textbf{(d)} Global style transfer (oil painting).}
    \label{fig:trigger_examples}
\end{figure*}

To provide a concrete understanding of the threat models, we visualize the poisoned samples (triggers) used in our experiments in Figure~\ref{fig:trigger_examples}.

%%%%%%%%%%%%%%%%%%%%%%%%%%%%%%%%%%%%%%%%%%%%%%%%%%%%%%%%%%%%%%%%%%%%%%%%%%%%%%%
%%%%%%%%%%%%%%%%%%%%%%%%%%%%%%%%%%%%%%%%%%%%%%%%%%%%%%%%%%%%%%%%%%%%%%%%%%%%%%%

\end{document}